\documentclass[prl,aps,twocolumn,preprintnumbers]{revtex4}
\usepackage{epsfig,amssymb,amsfonts}

\begin{document}
\title{Quantum Gates Between Two Spins in a Triple Dot System with an Empty Dot}
\date{\today}
    \author{Jose \surname{Garcia Coello}}
    \email{jose@theory.phys.ucl.ac.uk}
\author{Sougato Bose}
  \address{$^{1}$Department of Physics and Astronomy, University
College London, Gower St., London WC1E 6BT, UK}



\begin{abstract}
We propose a scheme for implementing quantum gates and entanglement between spin qubits in the outer dots of a triple-dot system with an empty
central dot. The voltage applied to the central dot can be tuned to realize the gate. Our scheme exemplifies the possibility of quantum gates
outside the regime where each dot has an electron, so that spin-spin exchange interaction is not the only relevant mechanism.  Analytic
treatment is possible by mapping the problem to a t-J model. The fidelity of the entangling quantum gate between the spins is analyzed in the
presence of decoherence stemming from a bath of nuclear spins, as well as from charge fluctuations. Our scheme provides an avenue for extending
the scope of two qubit gate experiments to triple-dots, while requiring minimal control, namely that of the potential of a single dot, and may
enhance the qubit separation to ease differential addressability.
\end{abstract}

\maketitle
\bibliographystyle{apsrev}

\section{Introduction}

Quantum Dots (QDs) are regarded as a good system for the storage and manipulation of Quantum Information (QI). In these systems, the qubit could
be encoded, for example, in the spin of an electron \cite{Divincenzo-Loss,Petta,Hanson,Burkard,Engel,marcus} or the electronic charge
distribution \cite{Jefferson} or even the presence/absence of excitons \cite{Lovett-Nazir-Briggs}. Spin qubits are particularly important because of their long
decoherence times. The earliest proposals advocated the use of the spin of a single electron in a quantum dot as a qubit with quantum gates
being realized by tuning the exchange coupling  between two quantum dots \cite{Divincenzo-Loss}. However, the exchange interaction between dots is not the easiest parameter to control. For this reason, some early experiments
\cite{Petta} and recent proposals \cite{Burkard} have focussed on qubits encoded on two spins in double dot systems, where the control parameter
is the energy mismatch between the quantum dots. This is motivated by the fact that the energy mismatch between dots can be simple to control, for example, through source-drain bias \cite{source-drain} or local electrostatic gates \cite{Petta}. It would thus be interesting
to have a protocol where one requires only the above control (namely the energy mismatch between dots) and is yet able to use a single spin as a
qubit. 
\begin{figure}

    \includegraphics[width=9cm]{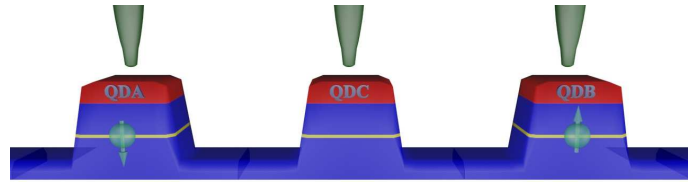}

\caption{The above figure depicts the triple dot system where we investigate the possibility of quantum gates. There are two spins in the
outer dots which behave as qubits, while the central dot is empty both before and after the quantum gates. QDA, QDC and QDB in the figure stand for quantum dots A, C and B respectively, while separate electrodes controlling the voltages of each dot are also shown in the figure.} \label{setup}
\end{figure}
In this paper, we propose such a protocol using a linear triple dot system where qubits (individual electronic spins) are placed in the
outer dots with the central dot being kept unfilled. An alternative motivation for our work stems from the fact that various triple dot systems
are now being fabricated and their charge stability diagram with small numbers of electrons is being studied \cite{tripledots,marcus3dot}. However,
 most experiments in quantum information context (with the exception of Ref.\cite{marcus3dot}) have so far have been limited exclusively to double dot systems. It would thereby be very timely to
have a scheme such as ours, which enhances the scope of quantum gate related experiments to triple dot systems. Of course, the most
straightforward generalization of the schemes in double dots \cite{Divincenzo-Loss} would be to have three spin qubits in three quantum dots
i.e., the filling of the quantum dots being $(1,1,1)$. Another possibility is to have a spin in the central dot as a {\em mediator} for an
effective coupling between the outer dots, a configuration which has recently been studied in the molecular context \cite{loss-nat}. Another possibility with a $(1,1,1)$ filling is to encode a single qubit in three dots \cite{DiVincenzo-Whaley}, which has been explored in a very recent experiment \cite{marcus3dot}. Here we
find out that a lower filling configuration, namely a $(1,0,1)$ filling, also provides a system for two qubit quantum gates with the qubits being in the
outer dots. The $(1,0,1)$ filling prevents one from reducing the problem to one of distinguishable spins (labeled by their sites) interacting
through exchange interactions as in the existing schemes for quantum gates with spin qubits. Thus both the tunneling of electrons from one site
to another, and careful second quantized treatment are important in the current problem and make it interesting. Note that very recently an alternative mechanism for two qubit gates in a triple dot system with qubits in the first two dots, i.e., a $(1,1,0)$ filling, has been proposed using spin dependent tunnelings and adiabatic processes \cite{kestner} -- however, that is a very different scheme from the one we report here which is neither adiabatic nor exploiting spin dependent couplings. Moreover, fast and coherent singlet-triplet filtering mechanisms have been proposed in single dots which effectively behave rather similarly to multiple quantum dots \cite{jefferson,coello}.


\section{setup}
 Our setup consists of 3 quantum dots (QDs) in a row, with the voltage applied to the central one being controllable by some
electrode, as shown in Fig.\ref{setup}. We label the outer dots of the chain as dot A and dot B, while we label the central dot as dot C. We
will assume that the Mott-Hubbard Hamiltonian describes the system well (for example, see Refs.\cite{stafford}), whereby the
relevant Hamiltonian is
\begin{eqnarray}
{\cal H}=\sum_{\sigma,i}E_i~d^{\dagger}_{i\sigma}d_{i\sigma}+\sum_{\sigma}t_{\text{AC}} (d^{\dagger}_{\text{A}\sigma}d_{\text{C}\sigma} +
d^{\dagger}_{\text{C}\sigma}d_{\text{A}\sigma})\nonumber\\+\sum_{\sigma}t_{\text{CB}} (d^{\dagger}_{\text{C}\sigma}d_{\text{B}\sigma} +
d^{\dagger}_{\text{B}\sigma}d_{\text{C}\sigma})+\frac{1}{2}\sum_{i}U_{i} n_{i} (n_{i}-1). \label{H}
\end{eqnarray}
In the above, $i$ stands for A, B and C, $d^{\dagger}_{i\sigma}$ creates and $d_{i\sigma}$ annihilates an electron at the $i$th dot in the spin
state $\sigma$ with energy $E_i$. Here we have assumed that the particles are created only in the lowest energy state at the site ($E_i$) and
the higher energy levels for a single electron are so well separated that they never become involved in the problem. $U_i$ is the Coulomb
repulsion in the QD $i$, $n_{i}=\sum_{\sigma}d^{\dagger}_{i\sigma}d_{i\sigma}$ is the total electron number operator of the $i$th dot and
$t_{\text{AC}}$ and $t_{\text{CB}}$ are tunnel matrix elements betweens dots. Here we have assumed that another term, often present in Hubbard
models for dot arrays, namely the inter-dot electrostatic interaction is zero.  Moreover, we have assumed that there exists no tunneling between
the non-neighboring dots, namely A and B. This should be a good approximation in serial triple dot systems \cite{tripledots} as $A$ and $B$ have a
high separation. Some relevant experimental values for $E_i$, $U_i$, $t_{\text{AC}}$ and $t_{\text{CB}}$ from recent experiments are given in
the table $1$ of Ref.\cite {marcus}, which will provide our guide for exploring feasibility issues. The dots at the two ends (i.e., QD A and QD
B) are each assumed to be filled up by a single electron as shown in Fig.\ref{setup}. These two electronic spins will be the two qubits in our
problem. As these qubits are identified by their sites, they can be referred to as qubit A and qubit B respectively. Of course, we should be
able to control when we want to enact a quantum gate between the aforementioned qubits, and for those intervals of time when we do not want any
gates, nothing should happen to the qubits (the state of the qubits, whatever they are, should remain intact). To ensure this, one has to ensure
that the qubits stably remain in a $(1,0,1)$ filling as shown in Fig.\ref{setup} and do not hop into QD C during this non-processing stage. This
is achieved by choosing an appropriate set of voltages applied to the triple dot system and there are quite a few experimental examples by now
in which the $(1,0,1)$ filling has already been realized. Typically, if the Hamiltonian ${\cal H}$ of Eq.(\ref{H}) is valid with
$t_{\text{AC}}\approx t_{\text{CB}}=t$, then one has to set the volgate applied to QD C to a lower value and the voltages of QDs A and B to a
higher equal value. Also we have to work with systems with $t <<|E_{\text C}-E_{\text A}|,|E_{\text C}-E_{\text B}|$ so that hopping is severely
suppressed. In this "non-processing" mode of our system, the system evolution effectively freezes. When one intends to accomplish a quantum
gate, one rapidly sets $E_{\text C}=E_{\text A}=E_{\text B}$ and a time evolution starts (this is true as long as the Hamiltonian ${\cal H}$
with $t_{\text{AC}}\approx t_{\text{CB}}=t$ is a good approximation of the triple dot system in consideration; in different experimental
realizations, the Hamiltonian may deviate differently from this, and then, for the processing mode, one has to apply that voltage which ensures
the electrostatic energy of the configurations $(1,0,1),(1,1,0)$ and $(0,1,1)$ to be equal). We will show that a two qubit ``entangling" quantum gate can
be obtained between the qubits by virtue of this evolution through Hamiltonian ${\cal H}$. Though during the time evolution, the electrons can
hop into the otherwise empty QD C, and indeed this is necessary for their spins to interact, at the end of a fixed period of evolution, one
electron is back in each of QD A and QD B. We will assume that single qubit gates on the spins in the outer dots can be trivially implemented by
using local fields, so that we are going to concentrate only on the demonstration of a two qubit entangling gate. The demonstration of the two qubit entangling gate is at the heart of demonstrating the viability of a system for universal quantum computation.

\section{The two qubit gate}

 The specific gate that we will demonstrate as enactable between the spins in the outer dots by means of their evolution through the Hamiltonian ${\cal H}$ is given by the
 following evolution of the computational basis states $|\uparrow\rangle$ (up spin along any axis, say $z$, standing for the logical state $|0\rangle$) and $|\downarrow\rangle$ (down spin along
 any axis, say $z$ standing for the logical state $|1\rangle$):

\begin{eqnarray}
|\uparrow\rangle_A|\uparrow\rangle_B &\rightarrow& |\uparrow\rangle_A|\uparrow\rangle_B \nonumber\\
|\uparrow\rangle_A|\downarrow\rangle_B &\rightarrow& e^{i\frac{\pi}{4}}\frac{1}{\sqrt{2}}(|\uparrow\rangle_A|\downarrow\rangle_B - i |\downarrow\rangle_A|\uparrow\rangle_B)\nonumber\\
|\downarrow\rangle_A|\uparrow\rangle_B &\rightarrow& e^{i\frac{\pi}{4}}\frac{1}{\sqrt{2}}(|\downarrow\rangle_A|\uparrow\rangle_B - i |\uparrow\rangle_A|\downarrow\rangle_B)\nonumber\\
|\downarrow\rangle_A|\downarrow\rangle_B &\rightarrow& |\downarrow\rangle_A|\downarrow\rangle_B. \label{gate}
\end{eqnarray}
Note that the above gate is manifestly an entangling quantum gate as it takes the initial states $|\uparrow\rangle_A|\downarrow\rangle_B$ and
$|\downarrow\rangle_A|\uparrow\rangle_B$ to entangled states. Thus the above gate suffices, in conjunction with local unitary operations on
qubits A and B, for universal quantum computation \cite{bremner}.

 Before proceeding further, we have to briefly clarify the notations that we will use. The gate presented above is in the usual notation of
states of multiple qubits, where all the qubits are distinguishable and each qubit has its own distinct label. However, this distinctive labels
(namely, qubit A and qubit B) are true only in the ``non-processing" phase, i.e., before and after the time evolution by $H$. The two electrons
may loose their site labels (namely A and B) during the evolution and thereby a fully second quantized treatment which automatically takes
account of the indistinguishability of the electrons is necessary. So, as basis states for writing down the Hamiltonian of the system, we shall
use the states $d^{\dagger}_{i\sigma}d^{\dagger}_{j\sigma'}|0\rangle$ with $i,j=\text{A},\text{B},\text{C}$ and
$\sigma,\sigma'=\uparrow,\downarrow$, where $|0\rangle$ is the state with all three dots empty, and evaluate the matrix elements of the
Hamiltonian $H$ in this basis.

Let us point out that the total spin component along any axis is conserved by $H$. Choosing an axis to be the $z$ axis, for example, and
remembering our $(1,0,1)$ initial filling, the problem becomes three independent problems for the total $z$ component of the spin in the three
sites $S_z$ being $+1$ ($\sum_i d^{\dagger}_{i\uparrow}d_{i\uparrow}=2$), $0$ ($\sum_i d^{\dagger}_{i\uparrow}d_{i\uparrow}=1$) or $-1$ ($\sum_i
d^{\dagger}_{i\uparrow}d_{i\uparrow}=0$). In the $S_z=+1$ sector, a complete basis comprises three states
$d^{\dagger}_{\text{A}\uparrow}d^{\dagger}_{\text{C}\uparrow}|0\rangle, d^{\dagger}_{\text{A}\uparrow}d^{\dagger}_{\text{B}\uparrow}|0\rangle$
and $d^{\dagger}_{\text{C}\uparrow}d^{\dagger}_{\text{B}\uparrow}|0\rangle$, in which the $3\times 3$ Hamiltonian is simply
\[ H_{S_z=+1}=\left( \begin{array}{ccc}
0 & t & t \\
t & 0 & 0 \\
t & 0 & 0 \end{array} \right)\] From the above Hamiltonian it is easy to see that if the system starts in the two qubit state
$|\uparrow\rangle_A|\uparrow\rangle_B$ (which actually means the state $d^{\dagger}_{\text{A}\uparrow}d^{\dagger}_{\text{B}\uparrow}|0\rangle$),
then at times $\tau_m=m\frac{2\pi}{\sqrt{2}t}$, where $m$ is an integer, the system comes back to its original state without any phase factor.
Thereby, if we halt the evolution at any of these instances of time (by suddenly setting the voltages to the non-processing mode), we will have
the $|\uparrow\rangle_A|\uparrow\rangle_B\rightarrow|\uparrow\rangle_A|\uparrow\rangle_B$ part of the quantum gate in Eq.(\ref{gate}) satisfied.
Exactly the same result holds for the $|\downarrow\rangle_A|\downarrow\rangle_B\rightarrow|\downarrow\rangle_A|\downarrow\rangle_B$ part of the
quantum gate, which evolves in the $S_z=-1$ sector with an identical Hamiltonian matrix. Therefore it remains to check whether there exist any
values of $m$ for which the remainder of the quantum gate of Eq.(\ref{gate}) happens at $\tau_m$. For that we have to look at the Hamiltonian in
the $S_z=0$ sector.

\section{The evolution in the $S_z=0$ sector and demonstration of the gate}

\begin{figure}

    \includegraphics[width=9cm]{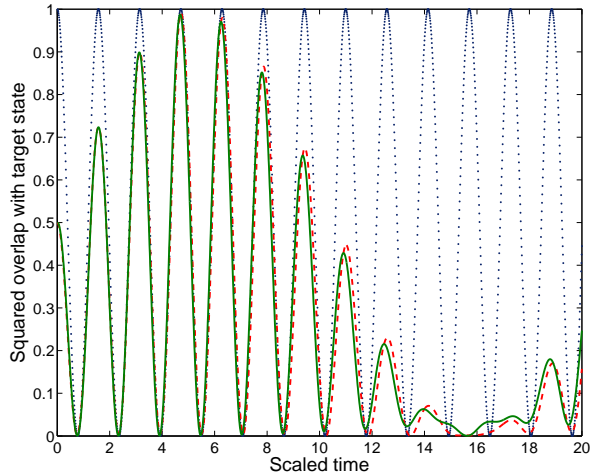}
\caption{Plots to demonstrate the occurrence of an entangling quantum gate at a certain instant of time between the spins A and B. The dotted line is the modulus squared overlap of $~~~~|\uparrow\rangle_A|\uparrow\rangle_B$ with the state it evolves to as a function of time after the gating Hamiltonian is switched on. Both the solid and the dashed lines show the modulus squared overlap of $~~~\frac{1}{\sqrt{2}}(|\uparrow\rangle_A|\downarrow\rangle_B - i |\downarrow\rangle_A|\uparrow\rangle_B)$ with the state to which $|\uparrow\rangle_A|\downarrow\rangle_B$ evolves as a function of time after the gating Hamiltonian is switched on. The solid line is from our analytic expression of Eq.(\ref{analyticexp}), while the dashed line from numerics without approximations. The parameters used in the plot are  $t=\sqrt{2}$ and $U=20$ in scaled units where the energy scale $10\mu$eV is set to unity (one unit of the scaled time
is about $0.1$ns.)} \label{pol1}
\end{figure}

In the $S_z=0$ sector a complete basis is made of the $9$ states $d^{\dagger}_{\text{A}\uparrow}d^{\dagger}_{\text{C}\downarrow}|0\rangle,
d^{\dagger}_{\text{A}\downarrow}d^{\dagger}_{\text{C}\uparrow}|0\rangle,
d^{\dagger}_{\text{C}\uparrow}d^{\dagger}_{\text{B}\downarrow}|0\rangle,
d^{\dagger}_{\text{C}\downarrow}d^{\dagger}_{\text{B}\uparrow}|0\rangle$,
$d^{\dagger}_{\text{A}\uparrow}d^{\dagger}_{\text{B}\downarrow}|0\rangle,
d^{\dagger}_{\text{A}\downarrow}d^{\dagger}_{\text{B}\uparrow}|0\rangle,
d^{\dagger}_{\text{A}\uparrow}d^{\dagger}_{\text{A}\downarrow}|0\rangle,
d^{\dagger}_{\text{B}\uparrow}d^{\dagger}_{\text{B}\downarrow}|0\rangle,
d^{\dagger}_{\text{C}\uparrow}d^{\dagger}_{\text{C}\downarrow}|0\rangle$. The $9\times 9$ Hamiltonian matrix in this basis is not reproduced
here for brevity, but it is important to note that here some elements such as $\langle 0|d_{\text{A}\uparrow}d_{\text{C}\downarrow} {\cal
H}d^{\dagger}_{\text{A}\uparrow}d^{\dagger}_{\text{A}\downarrow}|0\rangle$ are $t$, while others such as $\langle 0|
d_{\text{A}\downarrow}d_{\text{C}\uparrow} {\cal H}d^{\dagger}_{\text{A}\uparrow}d^{\dagger}_{\text{A}\downarrow}|0\rangle$ are $-t$. This sign
difference is important and cannot be obtained without proper second quantized treatment. Now assuming $U>>t$, one can adiabatically eliminate
the double occupancy states $d^{\dagger}_{\text{A}\uparrow}d^{\dagger}_{\text{A}\downarrow}|0\rangle,
d^{\dagger}_{\text{B}\uparrow}d^{\dagger}_{\text{B}\downarrow}|0\rangle,
d^{\dagger}_{\text{C}\uparrow}d^{\dagger}_{\text{C}\downarrow}|0\rangle$ to obtain the effective Hamiltonian

\[ {\cal H}_{eff}=\left( \begin{array}{cccccc}
-2J & 2J & -J & J & t & 0 \\
2J & -2J & J & -J & 0 & t \\
 -J & J & -2J & 2J & t & 0 \\
J & -J & 2J & -2J  & 0 & t \\
t & 0 & t & 0 & 0 & 0 \\
0 & t & 0 & t & 0 & 0 \end{array} \right)\] with $J=t^2/U$. The above effective Hamiltonian is that of a 3-site $t-J$ model, with parameter $t$
for hopping and parameter $J$ for a spin-spin interaction {\em only} when the spins are in neighboring sites. We define $\eta^\pm=-(3
J\pm\sqrt{9 J^2+2 t^2})$ and $\xi^\pm=\sqrt{2+(\eta^{\pm})^2/t^2}$, in terms of which, the eigenvalues of $H_{eff}$ are $\{0,-2 J,-\sqrt{2}
t,\sqrt{2} t,\eta^+,\eta^-\}$, while its eigenvectors are:

\begin{eqnarray}
|v_1\rangle&=&\{\frac{1}{2},\frac{1}{2},-\frac{1}{2},-\frac{1}{2},0,0\}\nonumber\\
|v_2\rangle&=&\{-\frac{1}{2},\frac{1}{2},\frac{1}{2},-\frac{1}{2},0,0\}\nonumber\\
|v_3\rangle&=&\{-\frac{1}{2 \sqrt{2}},-\frac{1}{2 \sqrt{2}},-\frac{1}{2 \sqrt{2}},-\frac{1}{2 \sqrt{2}},\frac{1}{2},\frac{1}{2}\}\nonumber\\
|v_4\rangle&=&\{\frac{1}{2 \sqrt{2}},\frac{1}{2 \sqrt{2}},\frac{1}{2 \sqrt{2}},\frac{1}{2 \sqrt{2}},\frac{1}{2},\frac{1}{2}\}\nonumber\\
|v_5\rangle&=&\{\frac{\eta^+}{2t\xi^+},-\frac{\eta^+}{2t\xi^+},\frac{\eta^+}{2t\xi^+},-\frac{\eta^+}{2t\xi^+},-\frac{1}{\xi^+},+\frac{1}{\xi^+}\}\nonumber\\
|v_6\rangle&=&\{\frac{\eta^-}{2t\xi^-},-\frac{\eta^-}{2t\xi^-},\frac{\eta^-}{2t\xi^-},-\frac{\eta^-}{2t\xi^-},\frac{1}{\xi^-},-\frac{1}{\xi^-}\}
\end{eqnarray}
We want to show that the initial state $|\uparrow\rangle_A|\downarrow\rangle_B$ of qubits A and B evolves to
$e^{i\pi/4}\frac{1}{\sqrt{2}}(|\uparrow\rangle_A|\downarrow\rangle_B - i
|\downarrow\rangle_A|\uparrow\rangle_B)$ at a certain time under the action of
the Hamiltonian ${\cal H}_{eff}$. Moreover this time must be coincident or approximately coincident with $\tau_m=m\frac{2\pi}{\sqrt{2}t}$ (discussed in the previous section) for
some $m$, so that the gate of Eq.(\ref{gate}) is accomplished at the time $\tau_m$. The initial state $|\uparrow\rangle_A|\downarrow\rangle_B$,
or more accurately the second quantized state $d^{\dagger}_{\text{A}\uparrow}d^{\dagger}_{\text{B}\downarrow}|0\rangle$, evolves with time
$\tau$ as:
\begin{eqnarray}
|\psi_{A\uparrow,
B\downarrow}(\tau)\rangle&=&\frac{1}{2}\{e^{i\sqrt{2}t\tau}|v_3\rangle+e^{-i\sqrt{2}t\tau}|v_4\rangle\}\nonumber\\&-&\frac{e^{-i\eta^+\tau}}{\xi^+}|v_5\rangle+\frac{e^{-i\eta^-\tau}}{\xi^-}|v_6\rangle
\end{eqnarray}
If we now once more invoke $U>>t$ to neglect terms of $O(t/U)$, we can simplify the mod-squared overlap of
$|\psi_{A\uparrow,B\downarrow}(\tau)\rangle$ with the target  $\frac{e^{i\pi/4}}{\sqrt{2}}(|\uparrow\rangle_A|\downarrow\rangle_B -i
|\downarrow\rangle_A|\uparrow\rangle_B)$ to the analytic expression
\begin{eqnarray}
&&\frac{\cos^2{\sqrt{2}t\tau}}{8}[\{1+\sqrt{2}\cos{(3J\tau-\frac{\pi}{4})}\}^2 \nonumber\\&+&\{1-\sqrt{2}\cos{(3J\tau+\frac{\pi}{4})}\}^2].
\label{analyticexp}
\end{eqnarray}
Notice that there are two distinct frequencies in the above expression, namely the higher frequency  $\sqrt{2}t$, which is due to the tunneling,
and the much lower frequency $3J$, which is due to the spin-spin interactions. Also note that, as expected, the modulus squared overlap with the
target state is $0.5$ at time $\tau=0$. However, most important to note is that at times $\tau^{'}_n=(2n+1)\pi/6J$ with $n$ being an integer,
the modulus squared overlap is unity implying that at these instances, the initial state $|\uparrow\rangle_A|\downarrow\rangle_B$ of qubits A
and B has fully evolved to the entangled state $e^{i\frac{\pi}{4}}\frac{1}{\sqrt{2}}(|\uparrow\rangle_A|\downarrow\rangle_B - i
|\downarrow\rangle_A|\uparrow\rangle_B)$. By following identical steps as above, one can prove that at times $\tau^{'}_n$ the initial state
$|\downarrow\rangle_A|\uparrow\rangle_B$ of qubits A and B evolves to $e^{i\frac{\pi}{4}}\frac{1}{\sqrt{2}}(|\downarrow\rangle_A|\uparrow\rangle_B - i
|\uparrow\rangle_A|\downarrow\rangle_B)$. As $2\pi/\sqrt{2}t<<\pi/6J$, for any $\tau^{'}_n$ there will exist several values of $m$ for which
$\tau_m$ is close to $\tau^{'}_n$. Thus one can always choose some $m$ and $n$ so that $\tau_m\approx\tau^{'}_n$ and at this particular time the
quantum gate of Eq.(\ref{gate}) is accomplished. Ideally we would like to choose the shortest possible time to accomplish the quantum gate to
minimize the effects of decoherence. The earliest opportunity is at time $\tau^{'}_0$ as this is the earliest time the second and third lines of
the gate of Eq.(\ref{gate}) is accomplished. Depending on the strength of the tunnel coupling $t$, nearly always it is possible to find a $m$
such that $\tau_m\approx\tau^{'}_0$ so that the quantum gate of Eq.(\ref{gate}) is accomplished at $\tau^{'}_0$. To convince the readers about
this, we take explicit values of parameters in scaled units. First we set the energy scale of about $10\mu$eV, which is a realistic typical scale of $t$ \cite{marcus,hawrylak,yamamoto}, to unity. In these units, we take $t=\sqrt{2}$ and $U=20$, so that $U>>t$ is valid and yet $J\sim 0.1$ is not too small. Such ratios of
$U/t$ are available and realistic \cite{hawrylak,yamamoto}), and based on them we plot some relevant curves in Fig.\ref{pol1}.

It is clear from the
figure that the modulus squared overlaps of the $|\uparrow\rangle_A|\uparrow\rangle_B$ state with itself and the
$|\uparrow\rangle_A|\downarrow\rangle_B$ with $\frac{1}{\sqrt{2}}(|\uparrow\rangle_A|\downarrow\rangle_B - i
|\downarrow\rangle_A|\uparrow\rangle_B)$, both achieve values indistinguishable from unity at time $\tau^{'}_0$. Further note that if one could
always tune the two free parameters $t$ and $U$, to ensure that $\tau^{'}_0\approx\tau_m$ holds for some $m$. Fig.\ref{pol1} also
presents a plot for the evolution of $|\uparrow\rangle_A|\downarrow\rangle_B$ to $\frac{1}{\sqrt{2}}(|\uparrow\rangle_A|\downarrow\rangle_B - i
|\downarrow\rangle_A|\uparrow\rangle_B)$ from exact numerical diagonalization of Eq.(\ref{H}) to show that the approximations (adiabatic
elimination) leading to the expression of Eq.(\ref{analyticexp}) is valid. However, to verify the quantum gate, one also needs to verify the phases outside the brackets on the right hand sides of the second and third lines of Eq.(\ref{gate}). We temporarily postpone this, and will verify these through additional plots that we make in the next section where we treat decoherence.

\section{Role of noise and decoherence}

 Now that we have demonstrated the possibility of an entangling gate between the spin qubits in our triple dot setting, we proceed to
investigate how this gate is affected by various sources of decoherence. During the fleetingly small time window
of gate operation (about a nanosecond) transient charge superpositions will exist, and thereby the gate will be subject to some charge decoherence despite operating between spin qubits. Note that this is {\em not unique} to our setting, but, in fact, also automatically present when one intends to implement two qubit gates with singlet-triplet qubits defined in double dots. There the singlet and the triplet have to go to distinct charge configurations to enable gates between two double-dot qubits \cite{Burkard}. As such decoherence is only during the gate operation, one can suppress it effectively by making the gate faster (i.e., $J$ stronger). In our case, during storage of the qubits, though, only spin decoherence, primarily due to the hyperfine interaction with nuclear spins, will be present.

\begin{figure}

    \includegraphics[width=9cm]{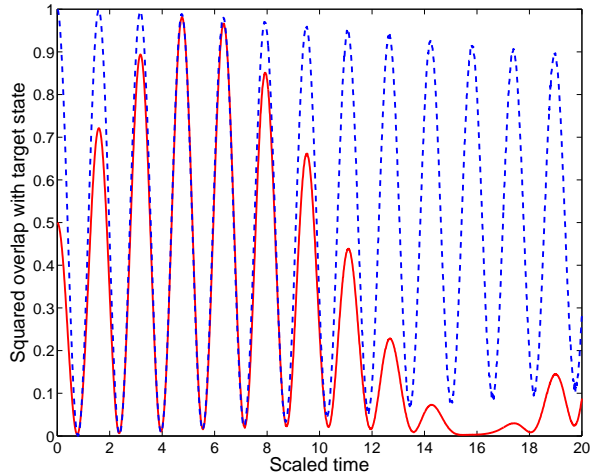}

\caption{The figure shows the effect of charge decoherence on the quantum gate of our protocol. We induce a charge decoherence time-scale of about 1 ns (about 10 units of our scaled time) by appropriately tuning a $1/f$ noise. The time evolution of the modulus squared overlap of an initial $|\uparrow\rangle_A|\uparrow\rangle_B$ state under this noise with itself (dashed curve) shows the purely charge based decoherence effect. Keeping the parameters of the charge noise the same, we have also plotted the modulus squared overlap of the state $\frac{1}{\sqrt{2}}(|\uparrow\rangle_A|\downarrow\rangle_B - i |\downarrow\rangle_A|\uparrow\rangle_B)$ with the state to which $|\uparrow\rangle_A|\downarrow\rangle_B$ evolves as a function of time after the gating Hamiltonian is switched on (solid curve).} \label{chargedec}
\end{figure}

We first model the effect of charge decoherence numerically. As the temperature is lowered
 enough so that the effect of phonons is eliminated (this assumption is met in current quantum dot experiments), decoherence due to spin-orbit interactions is suppressed. The $1/f$
noise generated in the triple dot device due to the fluctuations in the background charge is then the predominant
source of decoherence. We will phenomenologically fix the amplitude of this noise to set a charge decoherence time-scale of about $1$~ns (coherent charge oscillations have been observed till about $2$~ns \cite{shinkai} and even much higher have been reported in non-gated devices \cite{gorman}). Setting the amplitude in this phenomenological way
also has the advantage that it models charge decoherence of the best observed strengths irrespective of its cause (for example, some phonons may still be present). We have numerically generated a $1/f$ noise and used a distinct value of the noise in each time step. The numerical program that
 generates the noise guarantees that it has $1/f$ noise spectrum. We have also taken the tunneling $t$ to change with the mismatch of the dot energies -- we have taken $t$ to vary with the energy mismatch with a narrow gaussian profile of width $0.01$ (this profile of $t$ has been taken only for this phenomenological decoherence estimation and not elsewhere in the paper). We then vary the average strength of the fluctuations 
 till we get about a nanosecond time-scale of decay of the oscillations of the state $|\uparrow\rangle_A|\uparrow\rangle_B$ during the gate, which are essentially purely charge oscillations. This is plotted in Fig.\ref{chargedec}. We now take the {\em same} strength of noise for the evolution of $|\uparrow\rangle_A|\downarrow\rangle_B$ under the gate and numerically plot (in Fig.\ref{chargedec}) the probability of it to evolve to its ideal target state $e^{i\frac{\pi}{4}}\frac{1}{\sqrt{2}}(|\uparrow\rangle_A|\downarrow\rangle_B - i
|\downarrow\rangle_A|\uparrow\rangle_B)$. From the plot one can see that the effect of charge decoherence is not significant (the probability of the gate driving the initial states to their right targets is higher than $0.95$ for both states). This has happened because we have chosen parameters carefully enough to get a $J$ which can give a gate faster than the currently known charge decoherence rates.

\begin{figure}

    \includegraphics[width=9cm]{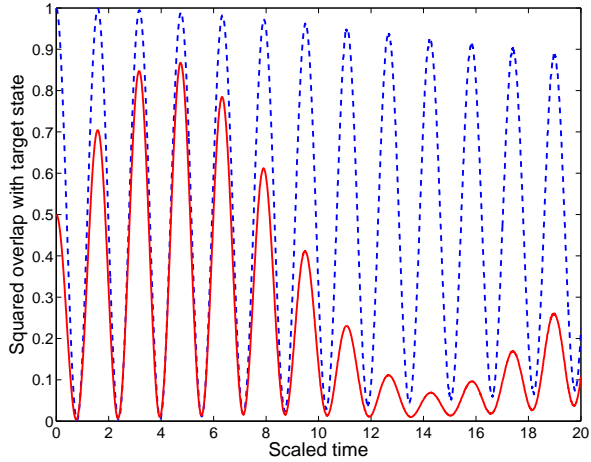}

\caption{This plot shows the combined effect of both hyperfine interactions and charge decoherence on the quantum gate proposed by us. Charge noise is set so as to have a charge decoherence time-scale of about 1 ns, while the strength of the random nuclear field causing the spin decoherence is set to the realistic value of $B_{\text{nuc}}\sim 0.1$ in scaled units (with $10 \mu$eV taken as unity). The time evolution of the modulus squared overlap of an initial $|\uparrow\rangle_A|\uparrow\rangle_B$ state under this noise with itself is shown as the dashed curve, while the modulus squared overlap of the state $\frac{1}{\sqrt{2}}(|\uparrow\rangle_A|\downarrow\rangle_B - i |\downarrow\rangle_A|\uparrow\rangle_B)$ with the state to which $|\uparrow\rangle_A|\downarrow\rangle_B$ evolves as a function of time after the gating Hamiltonian is switched on is shown as the solid curve.} \label{pol2}
\end{figure}

An additional form of decoherence that will be active is the nuclear baths in the quantum dots, which induce decoherence of the spin states. It is known that the orientations of the nuclear spins evolve at a much slower time-scale in comparison to the dynamics of the
 electrons (time-scales of $1/t$ and $1/J$) in quantum dot systems \cite{marcus} so that during one operation of our gate we may effectively regard the
 nuclear bath to provide a random but fixed (frozen in time) field. This is known as the quasistatic approximation \cite{marcus}. The effect of
 decoherence is then due to different constant fields in various runs of the gate (a distinct random direction and magnitude in each of the quantum dots for
 each run of the gate). Following the parameters given in Ref.\cite{marcus}, we have modeled the dynamics using a magnetic
 field of about an order of magnitude less than the tunneling $t$ in a random direction. The direction is chosen completely at random,
 while the magnitude is chosen from a Gaussian distribution given as $P(B)=\frac{1}{(2\pi B_{\text{nuc}}^2)^{3/2}}\exp{(-B^2/2B_{\text{nuc}}^2)}$. Here one cannot really use restricted
 spaces any more and the full Hilbert space of the problem is involved as the nuclear magnetic field connects these spaces. Thereby we tackle
 this part of the problem numerically in the full Hilbert space consisting of the $S_z=0,\pm 1$ sectors by exact diagonalization of ${\cal H}$
 with the
 addition of a random magnetic field term in each dot and using a charge decoherence of the same strength as before. The results are plotted in Fig.\ref{pol2} and show that the probability of successful occurrence of the quantum gate (Eq.(\ref{gate})) remains higher than $0.9$ for $B_{\text{nuc}}\sim 0.1$ in our units, which is comparable to its experimental values \cite{marcus}. In principle, though, this decoherence can be eliminated to a large degree by polarizing the background nuclear spins \cite{reiley} so that one can have quantum gates with fidelity only restricted by charge decoherence in a fleetingly small time window of gate operation. Even this latter decoherence should decrease with technology, and have already been reported to have very low values in non-gated devices \cite{gorman}. Alternatively it is known that quantum dot-like experiments can be performed also with {\em neutral} fermionic atoms in optical lattices \cite{Bloch} where charge decoherence is inactive.

\begin{figure}

    \includegraphics[width=9cm]{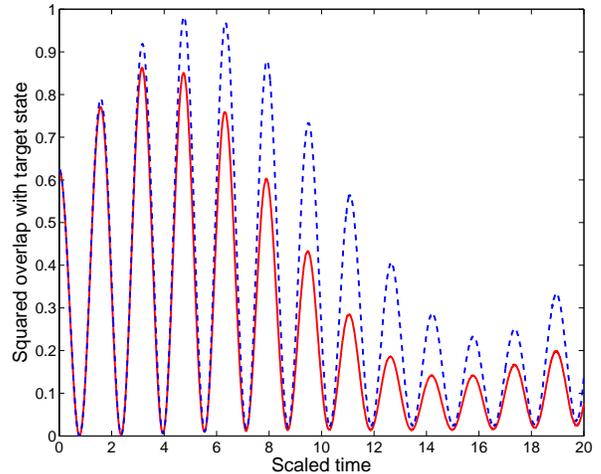}

\caption{This plot shows the effects of decoherence on an initial state $\frac{1}{\sqrt{2}}(|\uparrow\rangle_A|\uparrow\rangle_B+|\uparrow\rangle_A|\downarrow\rangle_B)$. It plots the evolution of the squared overlap of this state with is intended target state at the end of the gate, namely $\frac{1}{\sqrt{2}}|\uparrow\rangle_A|\uparrow\rangle_B+\frac{e^{i\pi/4}}{2\sqrt{2}}(|\uparrow\rangle_A|\downarrow\rangle_B - i |\downarrow\rangle_A|\uparrow\rangle_B)$. The dashed curve shows the evolution when only charge decoherence is present, while solid curve presents the evolution when both the charge as well as hyperfine induced decoherences are present. Charge noise is set so as to have a charge decoherence time-scale of about 1 ns, while the strength of the random nuclear field causing the spin decoherence is set to the realistic value of $B_{\text{nuc}}\sim 0.1$ in scaled units (with $10 \mu$eV taken as unity).} \label{pol3}
\end{figure}

Now we return to the issue of verifying all features of the gate of Eq.(\ref{gate}) through appropriate plots. For example, we need to verify that the phases outside the second and third lines of Eq.(\ref{gate}), and particularly, how it gets affected by decoherence. One way to examine this is to use $\frac{1}{\sqrt{2}}(|\uparrow\rangle_A|\uparrow\rangle_B+|\uparrow\rangle_A|\downarrow\rangle_B)$ as an initial state and verify how close it evolves to the ideal state (i.e., state under no decoherence) $\frac{1}{\sqrt{2}}|\uparrow\rangle_A|\uparrow\rangle_B+\frac{e^{i\pi/4}}{2\sqrt{2}}(|\uparrow\rangle_A|\downarrow\rangle_B - i |\downarrow\rangle_A|\uparrow\rangle_B)$ at time $\tau^{'}_0$. This is demonstrated under only charge decoherence and both charge and hyperfine interaction induced decoherences in Fig.\ref{pol3}.

\section{Gates in a high decoherence regime}
\begin{figure}

    \includegraphics[width=9cm]{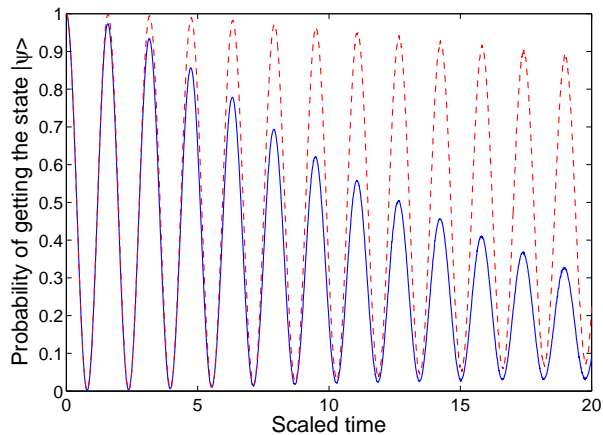}

\caption{This plot shows the effects of decoherence on an initial state $|\uparrow\rangle_A|\downarrow\rangle_B$. It plots the evolution of the squared overlap of this state, under both mechanisms of decoherence, with the state that it evolves to at any time $\tau$ under ideal conditions (i.e., $t<< U$ and no decoherence), namely $e^{i 3J\tau/2} (\cos{3J\tau/2}|\uparrow\rangle_A|\downarrow\rangle_B-i\sin{3J\tau/2}|\downarrow\rangle_A|\uparrow\rangle_B)$. The dashed curve shows the evolution when only charge decoherence is present, while solid curve presents the evolution when both the charge as well as hyperfine induced decoherences are present. Charge noise is set so as to have a charge decoherence time-scale of about 1 ns, while the strength of the random nuclear field causing the spin decoherence is set to the realistic value of $B_{\text{nuc}}\sim 0.1$ in scaled units (with $10 \mu$eV taken as unity).} \label{pol4}
\end{figure}

Suppose one has a very high charge decoherence (so that coherence stays, say, for only $0.1$~ns) then one can still use our triple-dot setup for a gate by stopping at the very first peak of the oscillation of the $|\uparrow\rangle_A|\uparrow\rangle_B$ state, i.e., at a time $\tau_1=2\pi/t\sim 0.1$~ns. The resulting quantum gate is however different and obtained by replacing the right hand sides of the second and third rows of Eq.(\ref{gate})  by $e^{i 3J\tau/2}~ (\cos{3J\tau/2}~|\uparrow\rangle_A|\downarrow\rangle_B-i\sin{3J\tau/2}~|\downarrow\rangle_A|\uparrow\rangle_B)$ and $e^{i 3J\tau/2}~ (\cos{3J\tau/2}~|\downarrow\rangle_A|\uparrow\rangle_B-i\sin{3J\tau/2}~|\uparrow\rangle_A|\downarrow\rangle_B)$ respectively (in the $t>>J$ limit). This has a lower entangling power, but is nonetheless an entangling gate, still useful for universal quantum computation. One merely has to halt the Hamiltonian at an earlier time (before decoherence has become too prominent) to get the gate and repeat the gate a few times to get a maximally entangling gate such as a CNOT from it. In Fig.\ref{pol4}, we have plotted the overlap of the ideal target state $e^{i 3J\tau/2} (\cos{3J\tau/2}|\uparrow\rangle_A|\downarrow\rangle_B-i\sin{3J\tau/2}|\downarrow\rangle_A|\uparrow\rangle_B)$ when one starts from the state $|\uparrow\rangle_A|\downarrow\rangle_B$ and has an evolution under the presence of both mechanisms of decoherence.

\section{Discussions} The primary achievement in this paper is to show that using triple dot systems, one can encode two single spin qubits and have an entangling quantum gate between them merely by tuning the voltage of the central dot (or voltage mis-alignment between the dots). This eases the restriction of having to tune the tunnel coupling $t$ on a fast time-scale, which might be difficult \cite{Burkard} or even impossible to tune in some setups of permanently built dots. One can scale this scheme to several qubits by using a one dimensional array in a $ABABAB...ABA$ scenario with the $A$ sites having single qubits and the $B$ sites being empty in the non-operative state of the system. Whenever a quantum gate between two qubits is required, we tune the voltage of only the $B$ site between the qubits to enable a gate between them. We have shown that the gate works with high enough fidelities for a variety of input states for achievable values of charge and spin decoherence rates. For stronger charge decoherence, one can halt the unitary evolution at earlier pertinent times and still get an entangling gate, albeit with lower power.

\section{Acknowledgements} SB acknowledges the support of the UK Engineering and Physical Sciences Research Council (EPSRC), the Royal Society and the Wolfson Foundation. JGC was supported by the EPSRC sponsored Quantum Information Processing Interdisciplinary Research Centre (QIPIRC). We have benefitted from discussions with John Jefferson, Brendon Lovett, Simon Benjamin and Andrew Briggs during early stages of this work. Additionally, during a visit to the Institute for Theoretical Atomic, Molecular and Optical Physics (ITAMP), Harvard, in 2008, SB  benefitted from some discussions with M. D. Lukin, J. M. Taylor and C. M. Marcus during the formative stages of this work.


%
%



\end{document}